\documentclass[11pt]{article}
\usepackage{colortbl}
\usepackage{natbib}
\usepackage{verbatim}
\usepackage{amssymb,amsfonts,amsmath,amsthm}
\usepackage{geometry}
\usepackage{alltt}
\usepackage{multirow}
\usepackage{graphicx}
\usepackage{subfigure}
\usepackage{setspace}
\usepackage{appendix}
\usepackage[colorlinks,bookmarksopen,bookmarksnumbered,citecolor=blue,urlcolor=green,hypertexnames=false]{hyperref}
\usepackage{dsfont}
\usepackage{geometry}
\bibpunct[ ]{(}{)}{;}{a}{}{,}
\onehalfspace
\begin{document}
\title{Fast estimation of multidimensional adaptive P-spline models\footnote{This paper is based on that published by the same authors as a part of the proceedings of the 30th International Workshop on Statistical Modelling, Linz, 6--10 July 2015 (Volume I, pp. 330\,--\,335. Eds: Friedl, H. and Wagner, H.)}}%
\author{Mar\'ia Xos\'e Rodr\'iguez - \'Alvarez$^{1,2}$, Mar\'ia Durb\'an$^{3}$, Dae-Jin Lee$^{1}$, Paul H. C. Eilers$^{4}$\\\\       
				\small{$^{1}$ BCAM - Basque Center for Applied Mathematics, Bilbao, Spain.}\\
				\href{mailto:mxrodriguez@bcamath.org}{mxrodriguez@bcamath.org}\\
				\small{$^{2}$ IKERBASQUE, Basque Foundation for Science, Bilbao, Spain}\\
        \small{$^{3}$ Department of Statistics, Universidad Carlos III de Madrid, Legan\'es, Spain}\\
        \small{$^{4}$ Erasmus University Medical Centre, Rotterdam, the Netherlands}}
\maketitle
\date{}
\sloppy
\begin{abstract}
A fast and stable algorithm for estimating multidimensional adaptive P-spline models is presented. We call it as \textit{Separation of Overlapping Penalties} (SOP) as it is an extension of the \textit{Separation of Anisotropic Penalties} (SAP) algorithm. SAP was originally derived for the estimation of the smoothing parameters of a multidimensional tensor product P-spline model with anisotropic penalties.
\end{abstract}
\section{Introduction}
Standard P-splines assume one smoothing parameter for modelling the effect of a covariate across its whole domain. In many applications it is desirable and needed to adapt smoothness locally to the data, and adaptive P-splines have been suggested.

The literature contains several proposals for adaptive P-splines. See, for instance, \cite{Krivobokova2008} and \cite{Wood2011}. However, the estimation procedures used by all these approaches can be very slow or even unstable. Based on the proposal by \cite{Wood2011}, we generalize the SAP algorithm given in \cite{Rodriguez2015} to deal with the adaptive penalty that is obtained. 

For brevity, we mainly focus here on the univariate adaptive approach for Gaussian data. However, the proposal can be extended to the multidimensional case, as well as to non-Gaussian responses along the lines of the Generalized Linear Model.
\section{Adaptive penalized splines}\label{MX:APS}
Consider a regression problem
\begin{equation*}
y_i = f\left(x_i\right) + \epsilon_i \;\;\;\;\; i = 1,\ldots n,
\end{equation*}
where $f$ is a smooth and unknown function and $\epsilon_i \sim N\left(0, \sigma^2\right)$. Within the P-spline framework \cite{Eilers1996}, the unknown function $f(x)$ can be approximated by a linear combination of B-splines basis functions, i.e., $f(x) = \sum_{j=1}^{c}\theta_{j}B_{j}\left(x\right)$, and smoothness is achieved by imposing a penalty on the regression coefficients $\boldsymbol{\theta}$ in the form 
\begin{equation}
P = \lambda\sum_{k = q + 1}^{c}\left(\Delta^q\theta_k\right)^2 = \lambda \boldsymbol{\theta}^{t}\boldsymbol{D}^{t}\boldsymbol{D}\boldsymbol{\theta},
\label{MX:1DPenalty}
\end{equation}
where $\lambda$ is the smoothing parameter, and $\Delta^q$ forms differences of order $q$ on adjancent coefficients, i.e., $\Delta\theta_k = \theta_k - \theta_{k-1}$, $\Delta^2\theta_k = \Delta\left(\Delta\theta_k\right) = \theta_k - \theta_{k-1} - \left(\theta_{k-1} - \theta_{k-2} \right) = \theta_k - 2\theta_{k-1} + \theta_{k-2}$, and so on for higher $q$. Finally, $\boldsymbol{D}$ is simply the matrix representation of $\Delta^q$. 

As can be observed in~(\ref{MX:1DPenalty}), the same smoothing parameter $\lambda$ applies to all coefficient differences, irrespective of their location. For more flexibility, we may think of assuming a different smoothing parameter for each difference
\begin{equation}
\sum_{k = q + 1}^{c}\lambda_{k-q}\left(\Delta^q\theta_k\right)^2 = \boldsymbol{\theta}^{t}\boldsymbol{D}^{t}\mbox{diag}(\boldsymbol{\lambda})\boldsymbol{D}\boldsymbol{\theta},
\label{MX:AdaptPenalty}
\end{equation}
where $\boldsymbol{\lambda} = \left(\lambda_1,\ldots,\lambda_{c-q}\right)^{t}$. Note that this approach would imply as many smoothing parameters as coefficients (minus $q$), which could lead to under-smoothing and unstable computations. Given the local and ordered nature of the coefficient differences, we may model the smoothing parameters $\lambda_k$ as a smooth function of $k$ (its position) and use B-splines for this purpose (here no penalty is assumed)
\begin{equation*}
\boldsymbol{\lambda} = \boldsymbol{C}\boldsymbol{\phi}, 
\end{equation*}
where $\boldsymbol{C}$ is a B-spline regression matrix of dimension $(c-q)\times p$ with $p < (c-q)$, and $\boldsymbol{\phi} = \left(\phi_1,\ldots,\phi_{p}\right)^{t}$ is the new vector of smoothing parameters. Performing some simple algebraic operations, it can thus be shown that the \textit{adaptive} penalty~(\ref{MX:AdaptPenalty}) is
\begin{equation}
\boldsymbol{\theta}^{t}\left(\sum_{l=1}^{p}\phi_{l}\boldsymbol{D}^{t} \mbox{diag}\left(\boldsymbol{c}_{l}\right)\boldsymbol{D}\right)\boldsymbol{\theta},
\label{MX:AdaptPenalty_smooth}
\end{equation}
where $\boldsymbol{c}_{l}$ denotes the column $l$ of $\boldsymbol{C}$.
\section{Estimation algorithm}
Estimation of the P-spline model subject to the penalty defined in~(\ref{MX:1DPenalty}) can be carried out based on the equivalence between P-splines and mixed models \cite{Currie2002}
\begin{equation*}
\boldsymbol{f} = \boldsymbol{B}\boldsymbol{\theta} = \boldsymbol{X}\boldsymbol{\beta} + \boldsymbol{Z}\boldsymbol{\alpha},
\end{equation*}
where $\boldsymbol{X}= \left[\boldsymbol{1}_n|\boldsymbol{x}|\ldots|\boldsymbol{x}^{(q-1)}\right]$ and $\boldsymbol{Z} = \boldsymbol{B}\boldsymbol{D}^{t}\left(\boldsymbol{D}\boldsymbol{D}^{t}\right)^{-1}$. It can be shown that the inverse of the variance-covariance matrix of the random effects $\boldsymbol{\alpha}$ is
\begin{equation*}
\boldsymbol{G}^{-1} = \frac{1}{\tau^2}\boldsymbol{I}_{c-q},
\end{equation*}
where $\tau^2 = \sigma^2/\lambda$. Here, only one variance component, $\tau$, is present, shrinking or penalizing the $\boldsymbol{\alpha}$ towards zero. The variance component can then be estimated on the basis of the iterative Harville-Schall (HS) algorithm \citep{Harville1977, Schall1991} . When applying the same mixed model parameterization to the adaptive P-spline with the penalty defined in~(\ref{MX:AdaptPenalty_smooth}), $\boldsymbol{G}^{-1}$ becomes
\begin{equation}
\boldsymbol{G}^{-1} = \sum_{l=1}^{p}\frac{1}{\tau_l^2}\mbox{diag}\left(\boldsymbol{c}_{l}\right),
\label{MX:Adaptive_G}
\end{equation}
where $\tau_l^2 = \sigma^2/\phi_l$. In this case each random effect is shrunk by several variance components, making the application of the HS algorithm unfeasible. In the paper by \cite{Rodriguez2015} the HS algorithm was extended to deal with multiple penalties on the same coefficients, with the penalties coming in that case from two (or more) spatial dimensions. However, multiple penalties can arise in a broader class of situations, as in our adaptive approach. Given that $\boldsymbol{G}^{-1}$ in ~(\ref{MX:Adaptive_G}) is expressed as a linear combination defined over the variance components, it can be shown that the SAP algorithm can be generalized to the estimation of $\tau_l$. Specifically, in each iteration the variance components estimates are updated, until convergence, according to
\begin{equation*}
\widehat{\tau}_l^2 = \frac{\boldsymbol{\alpha}^{t} \mbox{diag}\left(\boldsymbol{c}_{l}\right)\boldsymbol{\alpha}}{\mbox{ed}_l},
\end{equation*}
where
\begin{equation*}
\mbox{ed}_l = \mbox{trace}\left(\boldsymbol{Z}^{t}\boldsymbol{P}\boldsymbol{Z}\boldsymbol{G}\frac{\mbox{diag}\left(\boldsymbol{c}_{l}\right)}{\tau_l^2}\boldsymbol{G}\right),
\end{equation*}
with $\mathbf{P} = \boldsymbol{V}^{-1} - \boldsymbol{V^{-1}}\boldsymbol{X}\left(\boldsymbol{X^{t}V^{-1}X}\right)^{-1} \boldsymbol{X}^{t}\boldsymbol{V^{-1}}$ and $\mathbf{V} = \sigma^2\mathbf{I}_n + \mathbf{Z}\mathbf{G}\mathbf{Z}^{t}$. In principle, these traces involve several $n\times n$ matrices. However, an efficient computation can be achieved since: (a) $\boldsymbol{G}\mbox{diag}\left(\boldsymbol{c}_{l}\right)\boldsymbol{G}$ is a diagonal matrix; and, (b) $\boldsymbol{Z}^{t}\boldsymbol{P}\boldsymbol{Z}$ can be easily computed \citep[see equation (8) in][]{Rodriguez2015}. Finally, note that $\sum_{l=1}^{p}\mbox{ed}_l = \mbox{trace}\left(\boldsymbol{Z}^{t}\boldsymbol{P}\boldsymbol{Z}\boldsymbol{G}\right) = \mbox{trace}\left(\boldsymbol{Z}\boldsymbol{G}\boldsymbol{Z}^{t}\boldsymbol{P}\right)$, where $\boldsymbol{Z}\boldsymbol{G}\boldsymbol{Z}^{t}\boldsymbol{P}$ corresponds to the hat matrix of the random effects.

\section{Adaptive penalty in two dimensions}\label{MX:AdaptP_2D}
Extension of the univariate P-spline model given in Section~\ref{MX:APS} above to the modeling of two-dimensional (2D) surfaces is usually based on the tensor product of univariate B-spline basis, with the penalty matrix being defined as \cite[see][for further details]{Eilers2003}
\begin{equation*}
\gamma_{1}\left(\mathbf{I}_{c_2}\otimes\boldsymbol{D}_{1}\right)^{t}\left(\mathbf{I}_{c_2}\otimes\boldsymbol{D}_{1}\right) + \gamma_{2}\left(\boldsymbol{D}_{2}\otimes\mathbf{I}_{c_1}\right)^{t}\left(\boldsymbol{D}_{2}\otimes\mathbf{I}_{c_1}\right),
\label{MX:2DPenalty}
\end{equation*}
where $\gamma_1$ and $\gamma_2$ are the smoothing parameters (we assume anisotropy) and $\otimes$ denotes the Kronecker product. Following the same reasoning used for the univariate case, in the adaptive case each smoothing parameter $\gamma_{d}$ ($d = 1,\;2$) is replaced by a vector of smoothing parameters $\boldsymbol{\gamma}_{d}$, where each component is associated with one coefficient difference. To reduce the dimension, $\boldsymbol{\gamma}_{d}$ is then modeled by means of B-splines. However, since we still are in the two dimensional case, the tensor product of B-spline basis is used. Specifically,
\begin{align*}
\boldsymbol{\gamma}_{1} & = \left(\boldsymbol{C}_{11}\otimes\boldsymbol{C}_{12}\right)\boldsymbol{\phi}_1 = \boldsymbol{C}_{1}\boldsymbol{\phi}_1,\\
\boldsymbol{\gamma}_{2} & = \left(\boldsymbol{C}_{21}\otimes\boldsymbol{C}_{22}\right)\boldsymbol{\phi}_2 = \boldsymbol{C}_{2}\boldsymbol{\phi}_2,
\end{align*}
where $\boldsymbol{C}_{11}$, $\boldsymbol{C}_{12}$, $\boldsymbol{C}_{21}$ and $\boldsymbol{C}_{22}$ are B-spline regression matrices of dimension $(c_1 - q_1)\times p_{11}$, $c_2\times p_{12}$, $c_1\times p_{21}$ and $(c_2 - q_2)\times p_{22}$, respectively, and $\boldsymbol{\phi}_1 = \left(\phi_{11}, \cdots, \phi_{1p_{11}p_{12}}\right)$ and $\boldsymbol{\phi}_2 = \left(\phi_{12}, \cdots, \phi_{1p_{21}p_{22}}\right)$. Accordingly, the \textit{adaptive} penalty matrix in two dimensions can be then expressed as
\begin{align*}
& \sum_{m = 1}^{p_{11}p_{12}}\phi_{1m}\left(\mathbf{I}_{c_2}\otimes\boldsymbol{D}_{1}\right)^{t}diag\left(\boldsymbol{c}_{1,m}\right)\left(\mathbf{I}_{c_2}\otimes\boldsymbol{D}_{1}\right) + \\
& \sum_{s = 1}^{p_{21}p_{22}}\phi_{2s}\left(\boldsymbol{D}_{2}\otimes\mathbf{I}_{c_1}\right)^{t}diag\left(\boldsymbol{c}_{2,s}\right)\left(\boldsymbol{D}_{2}\otimes\mathbf{I}_{c_1}\right).
\end{align*}
$\boldsymbol{c}_{d,l}$ denotes the column $l$ of $\boldsymbol{C}_d$.
\section{Applications}
\begin{figure}[ht!]\centering
\includegraphics[width=15cm]{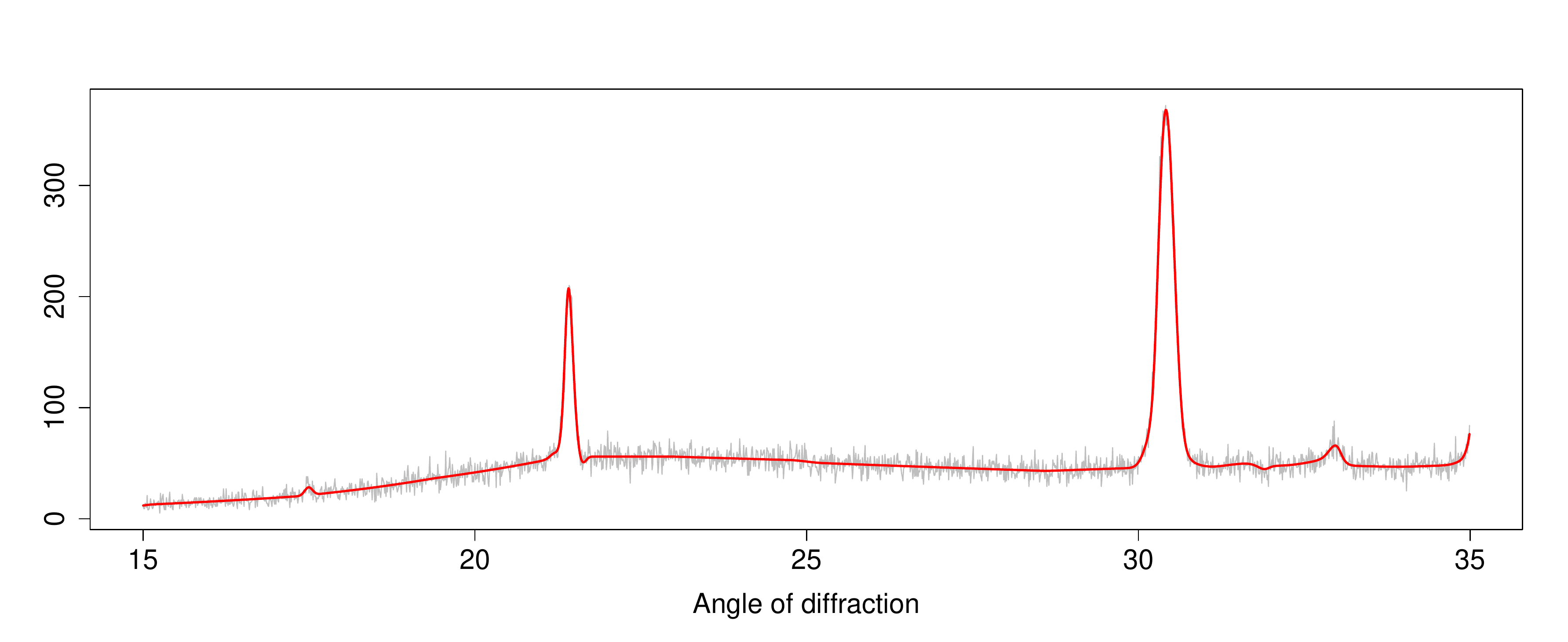}
\caption{\label{MX::Diffraction} Smooth effect of the angle of diffraction on the x-ray radiation. Grey: Raw data. Red: SOP algorithm.}
\end{figure}

To illustrate our proposal, we use data consisting of photon counts of diffracted x-ray radiation as a function of the angle of diffraction. The dataset can be found in the \texttt{R}-package \texttt{diffractometry} \citep{R2016, Davies2013}. Given that the outcome variable represents count data, a Poisson model was adopted. We compared the performance of the SOP algorithm with the method given in \cite{Wood2011}, as it is implemented in the \texttt{R}-package \texttt{mgcv}. In both cases, we used second-order differences and $200$ B-splines for the curve and $80$ for the adaptive penalty. Results are shown in Figure~\ref{MX::Diffraction}. The result of \texttt{mgcv} was almost identical to our proposal, so it is not depicted. Our algorithm took less that $3$ seconds, whereas \texttt{mgcv} was around $1000$ times slower. 

\begin{figure}[ht!]\centering
\includegraphics[width=15cm]{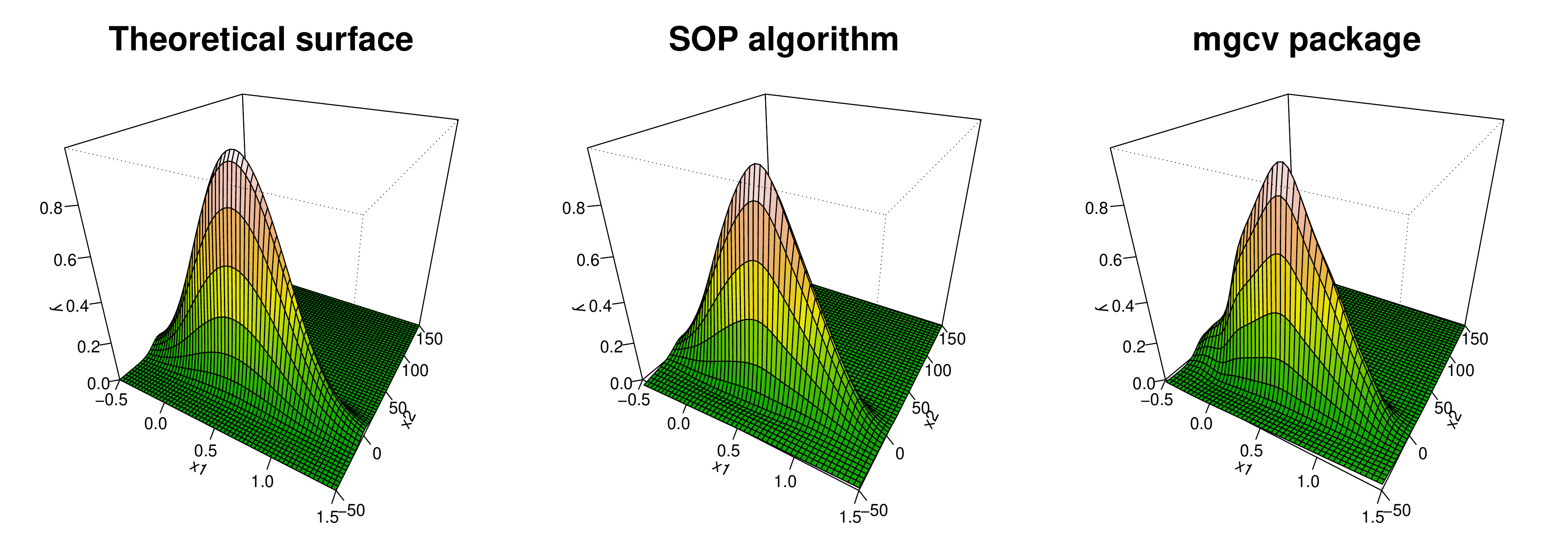}
\caption{\label{MX::2DAdaptive_Results} Simulated theoretical surface and fitted surface given by the SOP algorithm and the \texttt{mgcv} package.}
\end{figure}

\begin{figure}[ht!]\centering
\includegraphics[width=11.5cm]{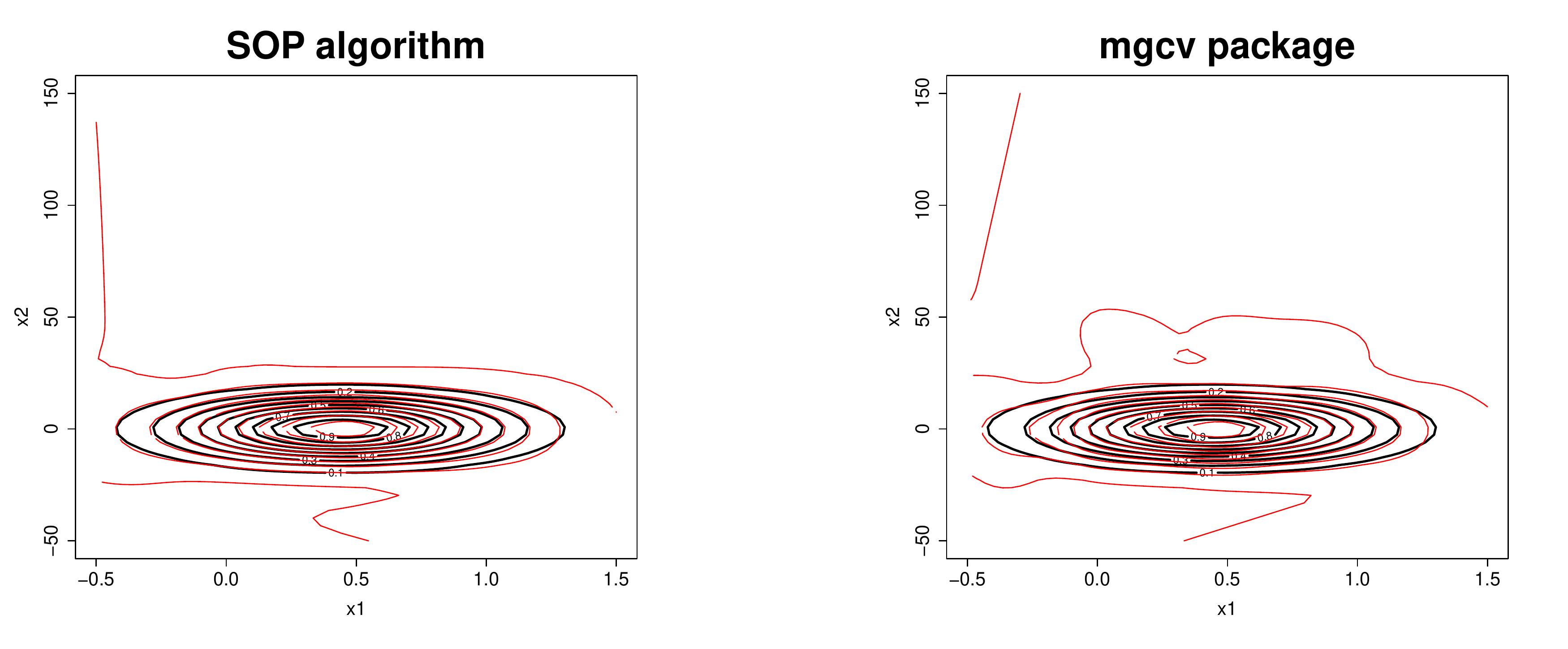}
\caption{\label{MX::2DAdaptive_Results_Contour} Contour plots. Black line: Simulated theoretical surface; Red line: fitted surface given by the SOP algorithm and
the \texttt{mgcv}package.}
\end{figure}

We also applied the proposed algorithm to the analysis of simulated 2D data. A sample size of $2000$ Gaussian data with $\sigma = 0.1$ was simulated, second-order differences were used, and we chose $15$ marginal B-splines for the surface and $8$ for the adaptive penalty. Note that this configuration yields to $128$ ($8\times 8\times 2$) smoothing parameters (or variance components). It should be noted, however, that the implementation of the 2D adaptive in the \texttt{mgcv} package uses a different adaptive penalty that the one presented on Section~\ref{MX:AdaptP_2D}. In that approach, and for the same configuration, the number of smoothing parameters is $64$. Here our algorithm was around $30$ times faster than \cite{Wood2011}'s approach, providing a computing time of about $22$ seconds. Figures~\ref{MX::2DAdaptive_Results}~and~\ref{MX::2DAdaptive_Results_Contour} depict the graphical results. The better behaviour of our approach can be explained by the fact that the adaptive penalty assumed in our case is more general and complex than that implemented in the \texttt{mgcv} package.

%
%

\section*{Acknowledgments}{The authors express their gratitude for the Spanish Ministry of Economy and Competitiveness MINECO grants MTM2014-55966-P, MTM2014-52184-P and BCAM Severo Ochoa excellence accreditation SEV-2013-0323, and for the Basque Government grant BERC 360 2014-2017.}

\end{document}